\documentclass[prx,twocolumn,superscriptaddress,floatfix,nopacs]{revtex4}
\usepackage{graphicx,amsfonts,amssymb,amsmath,hyperref,enumerate}

\newif\ifhyper
\hypertrue
\ifhyper
\hypersetup{
   citecolor = {green},
   colorlinks = {true}, 
   urlcolor = {blue} 
}
\fi

\newcommand{\beq}{\begin{equation}}
\newcommand{\eeq}{\end{equation}}
\newcommand{\beqa}{\begin{eqnarray}}
\newcommand{\eeqa}{\end{eqnarray}}
\newcommand{\ket} [1] {\vert #1 \rangle}

\def\ket#1{\vert#1\rangle}

\def\Longarrow{\protect\@lra}
\def\@lra{\relbar\joinrel\relbar\joinrel\relbar\joinrel%
          \relbar\joinrel\rightarrow}

\begin{document}

\title{Forecasting Election Polls with Spin Systems}

\author{Rub\'en Ibarrondo}
\affiliation{Department of Physical Chemistry, University of the Basque Country UPV/EHU, Apartado 644, 48080 Bilbao, Spain}

\author{Mikel Sanz}\email{mikel.sanz@ehu.es} 
\affiliation{Department of Physical Chemistry, University of the Basque Country UPV/EHU, Apartado 644, 48080 Bilbao, Spain}
\affiliation{Ikerbasque Foundation for Science, Maria Diaz de Haro 3, E-48013 Bilbao, Spain}
\affiliation{IQM, Nymphenburgerstr. 86, 80636 Munich, Germany}

\author{Rom\'an Or\'us}\email{roman.orus@dipc.org} 
\affiliation{Ikerbasque Foundation for Science, Maria Diaz de Haro 3, E-48013 Bilbao, Spain}
\affiliation{Donostia International Physics Center, Paseo Manuel de Lardizabal 4, E-20018 San Sebasti\'an, Spain}
\affiliation{Multiverse Computing, Paseo de Miram\'on  170,  E-20014  San  Sebasti\'an,  Spain}

\begin{abstract} 

We show that the problem of political forecasting, i.e, predicting the result of elections and referendums, can be mapped to finding the ground state configuration of a classical spin system. Depending on the required prediction, this spin system can be a combination of $XY$, Ising and vector Potts models, always with two-spin interactions, magnetic fields, and on arbitrary graphs. By reduction to the Ising model our result shows that political forecasting is formally an NP-Hard problem. Moreover, we show that the ground state search can be recasted as Higher-order and Quadratic Unconstrained Binary Optimization (HUBO / QUBO) Problems, which are the standard input of classical and quantum combinatorial optimization techniques. We prove the validity of our approach by performing a numerical experiment based on data gathered from \emph{Twitter} for a network of 10 people, finding good agreement between results from a poll and those predicted by our model. In general terms, our method can also be understood as a trend detection algorithm, particularly useful in the contexts of sentiment analysis and identification of fake news. 

\end{abstract}

\maketitle

\section{Introduction}
Political forecasting \cite{poli} aims at predicting the outcome of elections. As simple as it sounds, this problem is extremely complicated because of the huge number of factors which must be taken into account. Still, it is crystal-clear that this is an important problem, since bets on the succession of leaders are as old as human political organization. Nowadays, political parties spend millions in predicting the outcomes of elections via polls, extrapolations, regressions, and complex machine-learning techniques. Nevertheless, regardless of this game of thrones, the basic question here is actually more fundamental: can we somehow predict the social behaviour of humans? After all, people act both individually, driven by individual motivations, as well as collectively, driven by \emph{emergent} social trends. This question, unapproachable in practice for many centuries, has however taken a twist in our days because of easy access to petabytes of social data: likes and dislikes in \emph{Facebook}, retweets in \emph{Twitter}, GPS-location of cell phones... leaving alone the (undoubtedly fundamental!) issue of personal privacy for another discussion, our main question here is: can we predict human decisions employing a spin quantum model whose couplings are set based on their social behaviour as read from available data? Even though the question may sound shocking, it is noteworthy that the dynamics of groups in heavy metal concerts has been successfully modelled employing spin models \cite{SBSC13, ref1, ref2}. In fact, models in the context of sociophysics, such as mean field theory \cite{Maciej} and Galam models \cite{galam}, have already tried to provide an answer to this question for some time with relative success.  

In this article, we show that the above problem can be mathematically modelled by finding \emph{low-energy configurations of a spin system} and addressed employing quantum annealing. Individual voters are represented by spins, which point in different directions according to the individual's own political compass (i.e., the space of political alternatives). Social interactions and personal preferences shape decisions in the elections, and these can be precisely modelled by interacting coupling terms and magnetic fields in the corresponding classical spin Hamiltonian (energy cost function). In turn, the value of couplings and fields can be modelled from available social data. Our method admits a variety of interpretations. On the one hand, it can be seen as a tool for political forecasting. On the other hand, it can be understood more generically as an optimization algorithm for trend detection. This last point of view is particularly useful in the context of, e.g., sentiment analysis and identification of fake news. In this sense, our algorithm for solving such tasks is not based on machine learning or natural language processing, \emph{but rather on pure combinatorial optimization}, which is mostly unexplored in this context in the algorithmic sense. 

The structure of this paper is as follows. In Sec.\ref{compass} we explain the concept of political compass. Then, in Sec.\ref{spin} we show how to build the spin model, by defining first what we call ``political spin", and then ``political Hamiltonian". In this section, we also see how certain types of elections, such as referendums with binary questions, admit simple spin models which turn out to be NP-Hard. In Sec.\ref{sechubo}, we explain how the problem can be mapped into a Higher-order and Quadratic Unconstrained Binary Optimization (HUBO / QUBO) problem, which is the standard input for many optimization techniques. In Sec.\ref{experiment} we show the validity of our idea by performing an experiment on a network of 10 anonymous volunteers, who gave us their consent to use their social data from \emph{Twitter}, finding good agreement between the results of a poll and our predicted outcome. Finally, in Sec.\ref{conclusions} we wrap up our conclusions and future prospects. 

\section{Political compass}
\label{compass}

Let us begin by considering the notion of \emph{political compass}. In a nutshell, this is a configuration space where voters are placed according to their ideology. The space can have different axis, corresponding to different ideological aspects. For instance, one axis could be ``left -- right", another axis could be ``libertarian -- authoritarian", and so forth. The extreme points of these axis correspond to ideologies which are furthest away, e.g., extreme left vs extreme right in the ``left -- right" axis, with intermediate positions corresponding to intermediate ideologies. For actual examples of political compass, see Ref.\cite{compass1, compass2}. 

For the sake of simplicity, let us consider for the time being the case of one axis only, say, ``left -- right" (later on we will generalize our discussion to the many-axis case). Every voter $i$ has its own political compass (i.e., ideological configuration space), and his/her ideological preferences define a point on this compass. At the time of elections, the vote of individual $i$ will be according to his ``state" in the political compass at that time. This ``state" is influenced by many factors, such as likes and dislikes of statements by other individuals $j$, which are themselves also voters. It may also be influenced by geopolitical news, political scandals, economic decisions, and things alike. In the end, every individual will vote according to such factors. Some factors will have a positive influence in the individual (i.e., making him/her more likely to vote in favour of some ideology), whereas other factors will have a negative influence (i.e., making him/her more likely to vote against some ideology). 

\section{Spin model}
\label{spin}

\subsection{Political spin}
\begin{figure}
	\centering
	\includegraphics[width=0.55\linewidth]{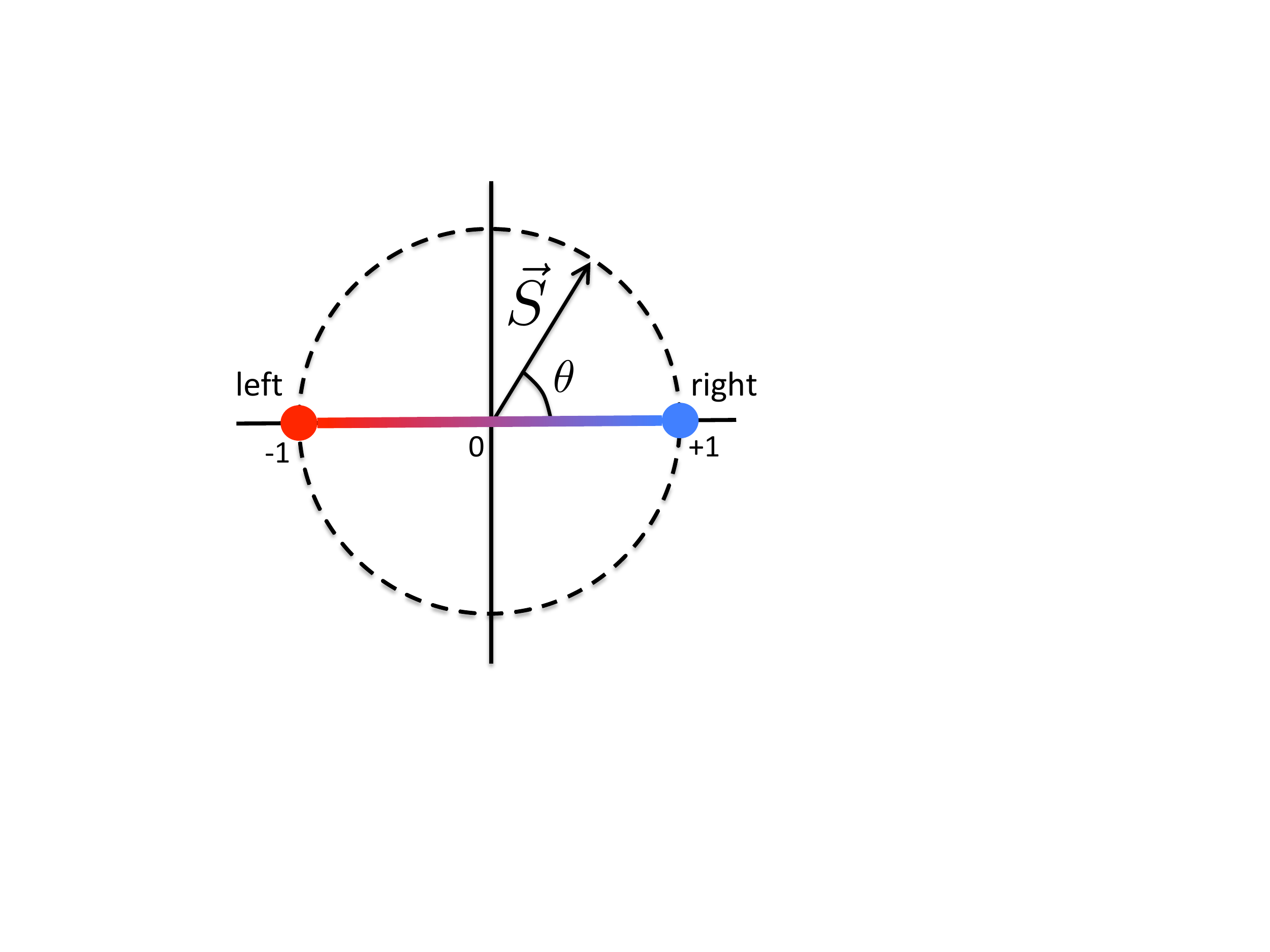}
	\caption{A one-axis ``left -- right" political compass, and the vector $\vec{S}$ as in Eq.(\ref{vector}) used to describe it.}
	\label{fig1}
\end{figure}

In what follows we show that the above can be accurately modelled by a classical spin system. Let us start by assigning a vector $\vec{S_i}$ to each voter $i$. This is a two-dimensional vector of unit length, i.e., 
\beq
\vec{S}_i = \cos \theta_i \hat{e}_1 + \sin \theta_i \hat{e}_2, 
\label{vector}
\eeq
with $\theta_i \in [0, 2\pi]$ and $\hat{e}_1, \hat{e}_2$ two orthonormal vectors. The vector is normalized to one ($\vec{S}^2_i = 1$), and angle $\theta_i$ parametrizes the position of voter $i$ in a one-axis political compass. More precisely, the position in the compass is the first component of the vector, $ \hat{e}_1 \cdot \vec{S}_i = \cos \theta_i$, with $\pm 1$ the extreme points, see Fig.\ref{fig1}. The vector $\vec{S_i}$ is nothing but a classical spin living on a two-dimensional plane. 

\subsection{Political Hamiltonian} Next, we define an energy (cost) function that is minimized by the ``most reasonable" outcome of the elections, understood as the configuration that globally satisfies as much as possible all the constraints mentioned above. This cost function will be nothing but a classical spin Hamiltonian, and will have different pieces. 

The first piece takes into account the \emph{influence environment} of each voter $i$. There will be positive influence, i.e., such that the voter will tend to vote in the same direction (in the compass) than some people, as well as negative influence, i.e., such that the voter will tend to vote in the opposite direction (in the compass) than some other people. These people are such that the voter likes (positive) or dislikes (negative) their opinions, and form the influence environment of the voter. Moreover, these people will be themselves voters $j$ with their own political compass parametrized by a vector $\vec{S}_j$. We thus model positive/negative influence via ferromagnetic/antiferromagnetic spin-spin interactions tending to align/anti-align vectors. Let us call $P_i$ and $N_i$ the sets of respectively all positive and negative influencers of voter $i$. We can then write the following classical interaction Hamiltonian $\mathcal{H}$ for all voters: 
\beq
\mathcal{H}_{int} = \sum_i \vec{S}_i \cdot \left( -\sum_{j \in P_i} \vec{S}_j + \sum_{j \in N_i} \vec{S}_j \right). 
\eeq
The above Hamiltonian is nothing but a two-body interacting classical $XY$ (compass) model with ferromagnetic and antiferromagnetic interactions. It is also not defined on any particular  lattice or graph: its interaction topology corresponds to that of \emph{correlated networks of influencers}. The correlations come from the fact that the sets $P_i$ and $N_i$ partially overlap amongst the different voters $i$.  In fact, any possible partial and asymmetric overlaps of the sets $P_i$ and $N_i$ for different voters $i$ is  possible, exactly as happens in real-life influence networks. 

The above Hamiltonian $\mathcal{H}_{int}$ takes into account only the positive and negative influence of the environment on a voter, but does not model possible penalties, rewards, preferences, and things alike. Luckily, all such terms can be easily introduced in the formalism by using local magnetic fields, tending to polarize spins along specific directions. For instance, we could consider a Hamiltonian of \emph{a priori preferences for each voter} such as
\beq
\mathcal{H}_{pre} = -\sum_i \vec{B}_i \cdot \vec{S}_i, 
\eeq
with $\vec{B}_i$ local polarising fields towards a specific direction, according to voter's $i$ a priori preferences. In fact, there will be voters for which the direction of the political compass will essentially be $100\%$ clear (say, people who are affiliated to political parties, influencers who publicly announce their preference, and so forth). Such voters have a fully-polarized spin direction, i.e. $|\vec{B}_i | = \infty$, which in practice means that their spin vector $\vec{S}_i$ is \emph{fixed}. Thus, for such voters, $\vec{S}_i$ is no longer a variable but \emph{a fixed parameter in the system.} These voters play the role of \emph{boundary conditions} in our spin Hamiltonian, reducing the degeneracy of configurations. 

Moreover, we can similarly add penalty or rewarding Hamiltonian terms for the different political parties (say, the different directions in the compass) depending on whatever relevant factors. For instance, if a given party $x$ has a corruption scandal, a global penalty term could be added such as 
\beq
\mathcal{H}_{pen} = \vec{\lambda}_x \cdot \left( \sum_i \vec{S}_i \right),  
\eeq
with $\vec{\lambda}_x > 0$ some penalty strength vector for party $x$. Rewarding terms can be similarly added. For instance, if party $y$ proposes to implement some popular measure, one may consider adding a term that favours its ideology by lowering the energy of their direction in the political compass, i.e., 
\beq
\mathcal{H}_{rew} = -\vec{\mu}_y \cdot \left( \sum_i \vec{S}_i \right), 
\eeq
with $\vec{\mu}_y > 0$ some reward strength vector for party $y$. Of course this is a simplified scenario, but one can easily see how to proceed in order to introduce further complications in the Hamiltonian. For instance, penalties and rewards could also be voter-dependent, and so forth. 

What we have shown above is that, at the end of the day, all the information available can in principle be codified in a cost function that corresponds to the Hamiltonian of a classical $XY$ model with spin-spin ferromagnetic and antiferromagnetic interactions on an arbitrary graph, together with local and global magnetic fields, and with some fixed spins as boundary conditions. In the case we are discussing, this is given by  
\beq
\mathcal{H} = \mathcal{H}_{int} + \mathcal{H}_{pre} + \mathcal{H}_{pen} + \mathcal{H}_{rew}, 
\label{ham}
\eeq
where we used the fact that $\vec{a} \cdot \vec{b} = \vec{b} \cdot \vec{a}$ for real vectors $\vec{a}$ and $\vec{b}$. This expression can of course be generalised and tailored in infinitely-many different ways, depending on the particulars of each situation, but it will \emph{always} have the same structure: spin-spin interactions and magnetic fields. The ground state of the Hamiltonian in Eq.(\ref{ham}), i.e., its minimal energy configuration, corresponds to the ``most reasonable" outcome of the elections, defined as the configuration that maximally satisfies the imposed constraints. Vectors $\vec{S}_i$ in this ground-state configuration provide the direction of voter's $i$ political compass, in turn telling us who is  voter $i$ more likely to vote for. 

\subsection{Multidimensional generalization} 

In the case discussed so far we only considered a one-dimensional political compass. It is, however, easy to consider also the case of a multidimensional compass, with several axis corresponding to different ``ideology dimensions" (say, ``left -- right", ``libertarian -- authoritarian", and so on). This can be done by simply considering a separate Hamiltonian for each one of the axis, with independent variables. Hence, there will be an independent spin vector $\vec{S}_i^\alpha$ as in Eq.(\ref{vector}) for each compass axis $\alpha$ of voter $i$. The overall Hamiltonian will then be 
\beq
\mathcal{H} = \sum_\alpha \mathcal{H}^\alpha, 
\label{alpha}
\eeq
with $\mathcal{H}^\alpha$ as in Eq.(\ref{ham}) for each axis $\alpha$. Moreover, in our scheme it is also possible to include the fact that, perhaps, the different axis $\alpha$ of a political compass may not be fully independent. For instance, libertarian ideology tends to be traditionally more left-wing, and things alike. In our formalism, this is nothing but an interaction of the type 
\beq
\mathcal{H}_{ax} = \sum_{i, \alpha < \beta} J^{\alpha \beta}_i \vec{S}_i^\alpha \cdot \vec{S}_i^\beta, 
\label{axax}
\eeq
with $J^{\alpha \beta}$ the interaction strength between axis $\alpha$ and $\beta$, which can be positive or negative depending on the type of bias that we wish to introduce, and the dot $(\cdot)$ being the scalar product of the two vectors, i.e.,  $\vec{S}_i^\alpha \cdot \vec{S}_i^\beta \equiv \cos \theta_i^\alpha \cos \theta_i^\beta + \sin \theta_i^\alpha \sin \theta_i^\beta = \cos( \theta_i^\alpha - \theta_i^\beta)$. The final Hamiltonian will then be 
\beq
\mathcal{H} = \sum_\alpha \mathcal{H}^\alpha + \mathcal{H}_{ax}. 
\label{alpha2}
\eeq
As discussed before, the ground state will provide the configuration of $\vec{S}_i^\alpha$ indicating the most reasonable vote of individual $i$ in the different axis $\alpha$ of the political compass, according to the constraints introduced \footnote{In fact, a useful extra axis that one could introduce is ``absentism -- not absentism", i.e., essentially whether the people will go to vote or not. This axis can be treated exactly in the same way as we described so far, and will provide very relevant information about the actual outcome. In fact, a carefully-constructed Hamiltonian could tell us about what the people would be more likely to vote or not, even in the case in which the people do not go to vote at all. Notice also that blank votes can be estimated from voters whose ``arrows" in the ground state turn out to fall far from the available voting options.}. 

\subsection{Discretization}

Several aspects of the political Hamiltonian above are worth of more in-depth discussion. To begin with, one could consider a political spin with discretized angles $\theta_n = 2 \pi n / q$, with $n = 0, 1, \cdots, q-1$. In such a case, one has that 
\beq
\vec{S}_i \cdot \vec{S}_j = \cos \left(\frac{2 \pi}{q} (n_i - n_j)\right), 
\eeq
so that the $XY$ Hamiltonian in Eq.(\ref{ham}) boils down to the classical $q$-state vector Potts model.  The case $q=2$ is particularly interesting: this is the case of a political axis with only two antagonistic options, e.g., ``brexit yes/no", ``independence yes/no", and things alike. One obtains
\beq
\vec{S}_i \cdot \vec{S}_j = s_i s_j,
\label{yesno}
\eeq
with $s_i = \pm 1$ Ising variables (and the same for $j$). Therefore, in this case Eq.(\ref{ham}) reduces to a classical Ising model with spin-spin interactions and magnetic fields. Importantly, finding the ground state of such a model is well-known to be NP-Hard \cite{barahona}. And, in turn, this  proves that \emph{the problem of political forecasting is itself also NP-Hard}, because it is, at least, as hard as finding the ground state of an arbitrary Ising classical spin model.

\section{HUBO and QUBO problems} 
\label{sechubo}

\subsection{Generalities} 

Finding the ground state of the classical Hamiltonian discussed above can in fact be rewritten as a Quadratic Unconstrained Binary Optimization (QUBO) problem. This is important, because the QUBO formula to be optimized (which amounts to a classical Ising model with spin-spin interactions and magnetic fields) is the natural input of many optimization techniques. QUBO problems can also be solved in quantum processors for instance via quantum annealing \cite{qann}, variational quantum eigensolvers \cite{vqe} and approximate quantum optimization algorithms \cite{aqoa}, which implies that the problem of forecasting elections can in principle be thrown into a quantum computer, with the idea in mind of finding better solutions to the problem than those achievable with classical algorithms. 

Let us consider again the case of a one-axis political compass. In our procedure, we will first map the Hamiltonian in Eq.(\ref{ham}) to a Higher-order Unconstrained Binary Optimization (HUBO) problem, with up to 4-bit terms. As such, the HUBO problem may already be a valid input for some optimization algorithms, both classical and quantum. Then, we will reduce the higher-order interactions down to quadratic terms by using ancillas and following the procedure from Ref.\cite{ancilla} (and also successfully used in Refs.\cite{crash, crashexp, DCLSS19,HLSCCWS20}).  

In order to write a HUBO formula for our Hamiltonian, we write the two components of the spin vector $\vec{S}_i$ in Eq.(\ref{vector}) using for instance a binary encoding in terms of $2n$ classical bits as 
\beq
\cos \theta_i \approx \sum_{a= 1}^n x_{i,a} 2^{-a}, ~~~ \sin \theta_i \approx \sum_{a = 1}^n y_{i,a} 2^{-a} ,
\label{digi}
\eeq
with $x_{i,a}$ and $y_{i,a}$ the classical bit variables for individual $i$, and $a = 1, 2, \cdots, n$. In this way, the values of both trigonometric functions are digitalized between $0$ and $1$, but a priori they do \emph{not} fulfil the constraint  $(\cos \theta_i)^2 + (\sin \theta_i)^2 = 1$. This can imposed by the penalty term   
\beq
C_i  = \left( \left( \sum_{a= 1}^n x_{i,a} 2^{-a} \right)^2 + \left( \sum_{a= 1}^n y_{i,a} 2^{-a} \right)^2 -1 \right)^2, 
\eeq
which is minimized when the constraint is satisfied so that $C_i \approx 0$. Next, we can write a HUBO formula for this problem by rewriting the Hamiltonian from Eq.(\ref{ham}) in terms of the bit variables $x_{i,a}$ and $y_{i,a}$, and adding the constraints $C_i$ as penalty terms with large prefactor, so that the ground state belongs to the subspace where $C_i \approx 0$ for all $i$. The HUBO formula $\mathcal{H}^{HUBO}$ to minimize is thus given by 
\beq
\mathcal{H}^{HUBO} = \mathcal{H}(x_{i,a}, y_{i,a}) + \nu \sum_i C_i, 
\label{hubo}
\eeq
with $\nu > 1$. It is easy to see that, because of the structure of the spin-spin interactions, $\mathcal{H}(x_{i,a}, y_{i,a})$ contains at most 2-bit terms, whereas the $C_i$ contain up to 4-bit terms. This HUBO formula is already a valid input for optimization algorithms, classical and quantum, allowing for multi-bit (multi-qubit) interactions. For instance, a universal quantum processor could approximate the ground state of the ``quantized" version of such a Hamiltonian, simply by replacing classical bit variables $x_{i,a}$ by qubit operators $\hat{x}_{i,a}$ with eigenvalues $0,1$ and eigenstates $\ket{0}$ and $\ket{1}$ respectively, and similarly for $y_{i,a}$. This could be done explicitly using a variational quantum eigensolver \cite{vqe} and/or an approximate quantum optimization algorithm \cite{aqoa}.

The HUBO formula above could also be a valid input of a quantum annealing processor allowing for multiqubit interactions. However, state-of-the-art quantum annealers allow only for two-qubit interactions. Luckily enough, one can reduce the 4- and 3-bit terms from Eq.(\ref{hubo}) down to 2-bit terms at most (thus obtaining a QUBO problem) but at the cost of introducing ancillas, as explained in Refs.\cite{ancilla, crash, crashexp}. The basic idea of the approach is to recast the corresponding quantum Hamiltonian into a modified, effective Hamiltonian with 2-qubit interactions at most, and which has the same low-energy eigenspace. The modified Hamiltonian, however, uses ancillary qubits: one ancilla for every three-qubit term, and up to four ancillas for every four-qubit term. In this way, the HUBO problem can be reduced down to a QUBO problem, and is therefore amenable to current quantum annealers such as the D-Wave machine. We refer the reader to the afore-mentioned references for further details. 

Importantly, notice that there is a relevant exception to the case of a HUBO problem in what we explained so far: the case of a yes/no referendum. This is exactly the case that we discussed around Eq.(\ref{yesno}), i.e., when one has a one-axis political compass with only two possible antagonistic options. In such a case, we do not need to apply any digitalization as in Eq.(\ref{digi}), since the Hamiltonian is already a QUBO problem via the mapping $x_i \equiv (s_i + 1)/2$ between bit and classical spin variables. In such a case, the problem can be handled directly not only by universal quantum processors, but also by state-of-the-art quantum annealers, without the need to introduce any ancillary qubits. 

Last but not least, we stress that everything that we explained about HUBO and QUBO problems applies also quite straightforwardly to the case of a multi-axis political compass, even with axis-axis interactions as in Eq.(\ref{axax}). 

\subsection{Required resources}
Concerning resources, it is clear that the number of bits (or qubits) required depends on the specifics of the case being analyzed. But as an example, for the single-axis case of a yes/no referendum of $N$ voters, one needs exactly $N$ bits (or qubits). Also for $N$ voters, in the one-axis case and for the HUBO problem one would require $2Nn$ qubits, with $2n$ the number of bits in Eq.(\ref{digi}). If the HUBO problem is subsequently reduced to a QUBO problem, one then also needs to take into account the ancillas, and this will be directly proportional to the number of 3-bit and 4-bit terms in Eq.(\ref{hubo}). If, on top, one has a multi-axis compass, then one needs to repeat the same analysis for each axis, so that the number of bits or qubits will also be proportional to the number of axis in the political compass. Overall, for the HUBO problem we could say that the number of required qubits $n_{tot}^{HUBO}$ scales asymptotically as
\beq
n_{tot}^{HUBO} = O(2Nnm), 
\eeq
with $N$ the number of voters, $2n$ the number of bits in the digitalization, and $m$ the number of axis in the compass. 

Notice that, in practice, the number of voters $N$ may actually be very large (e.g., around 40 million for presidential elections in Spain), so that the number of required qubits may look like ridiculously insane in some real-life scenarios. Nevertheless, it should be possible to define the same type of Hamiltonian for \emph{renormalized spins}, essentially equivalent to \emph{clustering} in the jargon of data science. The new effective variables would then represent not individuals, but groups of individuals (for instance, the average behavior of all people living in the same neighborhood). Using such a renormalized description we loose the detail of the forecast but, however, we gain a lot of computational efficiency both in the description of the relevant interactions in the Hamiltonian as well as in the total number of required qubits. This is something that needs to be seriously taken into account in practical scenarios where the number of voters is way too large: \emph{cluster the behaviour of the voters, in order to gain computational efficiency}.  

It is interesting to analyze the possible performance of the proposed algorithms with current quantum processors. As of today, available quantum processors have a small number of qubits and are noisy and prone to errors: the so-called Noisy Intermediate-Scale Quantum (NISQ) devices. In our case, our algorithm adapts naturally to quantum hardware built for optimization, such as the quantum annealer processor of D-Wave. The latests version of this processor, called \emph{Advantage}, consists of $\approx 5000$ (incoherent) superconducting qubits connected via $\approx 35000$ couplers according to a \emph{Pegasus} topology. Morover, D-Wave has implemented a hybrid solution by combining the quantum processor together with classical pre- and post-processing, allowing to simulate very large systems with good performance. This system is able to solve problems with several thousands of fully-connected binary variables in a few hundreds of seconds, and keeps improving every day. With this in mind, according to the formula above for $n_{tot}$, for a yes/no referendum with $m=1$ and $n=1$ we could get results for a number of voters $N = O(10^3)$ in roughly hundreds of seconds. For larger voting populations, a clustering could be performed to reduce the number of relevant variables to thousands, which would indeed fit into the current quantum capabilities.  

\section{Experiment using \emph{Twitter} data} 
\label{experiment}

\subsection{Building $\mathcal{H}$ from social data} 
But, how do we actually build $\mathcal{H}$? It is clear that one should collect as much social data as possible about the individuals and about social tendencies, and then codify everything in the Hamitonian. This could be extracted from a variety of sources, such as social networks, internet traffic, GPS data, and so on. In today's society, technology allows to gather such information \emph{always as long as it is provided with the consent of the individuals} \footnote{Otherwise one may run into trouble, as in the scandal of \emph{Facebook} and \emph{Cambridge Analytica}.}. Useful data to construct the Hamiltonian can also be obtained from polls and historical tendencies, just to mention a couple of examples. Notice also that, since voting is usually secret, one should not have the exact result of the individual votes from previous elections as available data. One may have, however, average data for different city sectors (for instance). This is also useful data that can be entered as a bias in the Hamiltonian. Notice as well that the relevant data may change in time, say, people stop following somebody in \emph{Facebook} and start following somebody else in \emph{Twitter} (for instance). It is therefore very important to have the most up-to-date information in order to build accurate Hamiltonians for real-time predictions. 

\subsection{\emph{Twitter} and referendums}

To benchmark our ideas we implemented the following experiment. We took 10 anonymous volunteers and, with their consent, we gathered data from their \emph{Twitter} accounts. In particular, for each individual we harvested the latest 200 likes, retweets, and comments amongst themselves, and used them to build a model. For simplicity, we focused on interactions that are surely positive (such as likes and positive retweets). We denote as $a_{ij}$ the number of positive interactions that individual $i$ receives from individual $j$, i.e., likes and retweets made by $j$ of posts made by $i$. Notice that this quantity  is not symmetric, i.e., $a_{ij} \neq a_{ji}$. We then build the following model for referendums of the type  ``choose either A or B": 
\beq
\mathcal{H} = \sum_i h_i s_i - \sum_{ij} J_{ij} s_i s_j, 
\eeq
with $s_i = \pm 1$ a classical spin for individual $i$ ($+1$ for answer ``A" and $-1$ for ``B"), and where the local fields $h_i$ and couplings $J_{ij}$ are defined as 
\beq
J_{ij}  \equiv  -\frac{1}{2}\left(a_{ij} + a_{ji} \right) ~~~~~~ h_i  \equiv  f_i \sum_j a_{ij}. 
\eeq
In the above equations, $J_{ij}$ measures the strength of ``mutual likeness" (ferromagnetic), and $h_i$ the strength of individual opinions. Specifically, $f_i$ are polarization factors that play the role of our boundary conditions: $f_i = \pm 1$ if we know that individual $i$ will vote $s_i = \mp 1$ for sure, and $f_i = 0$ if we know nothing about the voting intentions of $i$. Individuals with large $|h_i| = \sum_j a_{ij}.$ are the \emph{influencers} of our network. In practice, it is important to know the voting intentions (and therefore $f_i$) of the influencers, since they may have a large impact in the final outcome (i.e., large polarizing field $h_i$).  

\subsection{Results}

\begin{figure}
	\centering
	\includegraphics[width=1\linewidth]{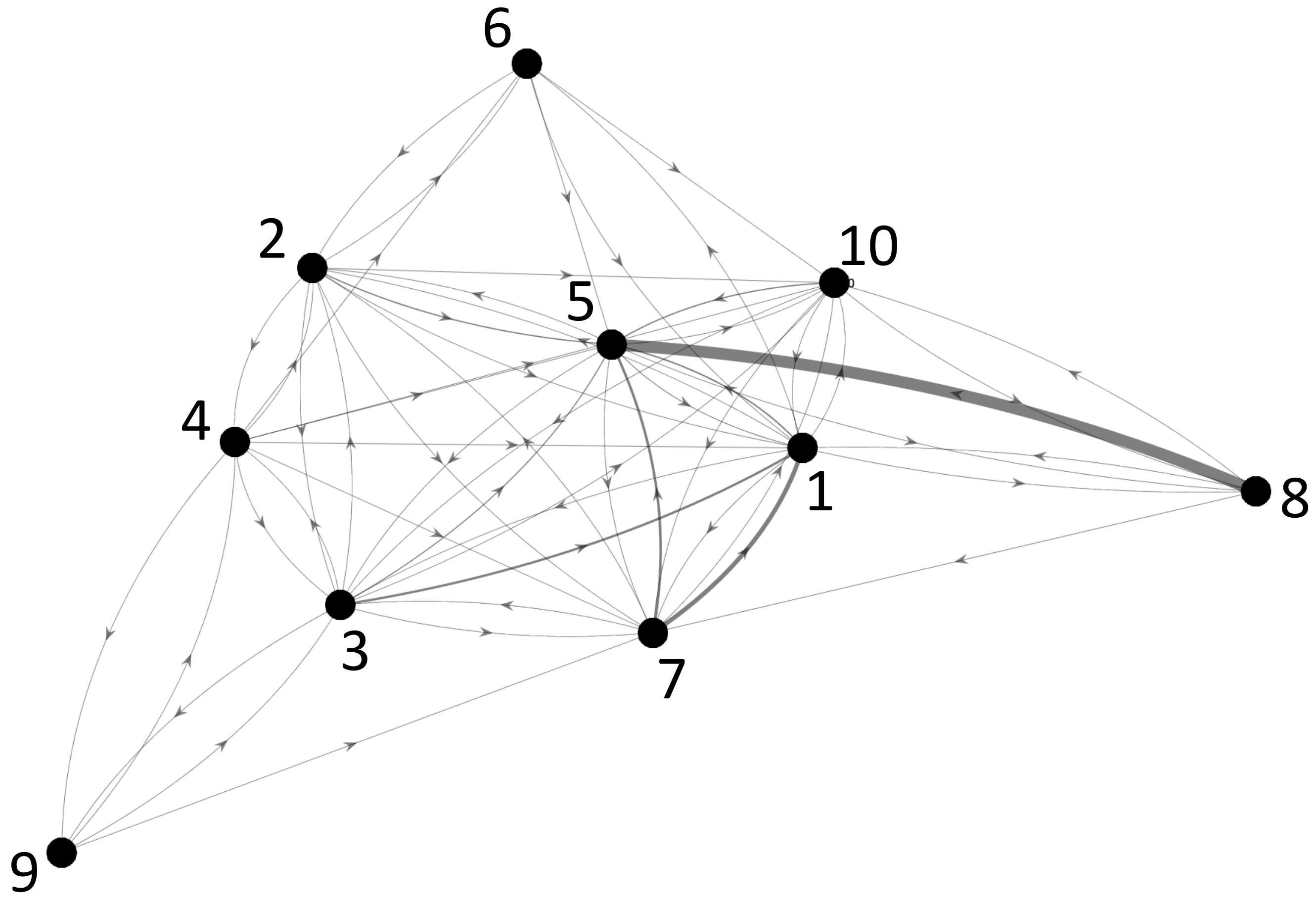}
	\caption{Graph corresponding to the social network of interactions for our experiment using \emph{Twitter} data. Nodes are individuals $i =1, 2, \cdots, 10$, and directed links correspond to interactions. The thickness of each link is proportional to the corresponding entry in matrix $a_{ij}$ from Eq.(\ref{matA}).}
	\label{graph}
\end{figure}

The interactions in our network are shown in the graph of Fig.(\ref{graph}). The matrix $a_{ij}$ that corresponds to the graph is given by 
\beq
a = 
\begin{pmatrix}
0   &  3 &    2 &    1 &    4  &  11 &   21 &    3 &    0  &   2 \\
     2   &  0 &    3 &    3 &    1 &    2 &    2  &   0  &   0  &   0 \\
     1  &   3  &   0  &   1  &   0  &   0  &   0  &   0  &   0  &   0 \\
     0  &   1  &   0  &   0  &   0  &   1 &    0 &    0 &    1 &    0 \\
     7  &   6  &   1  &   1  &   0  &   5  &  12  &  56  &   0  &   6 \\
     1  &   1  &   0  &   1  &   2  &   0  &   1  &   0  &   1  &   1 \\
     1  &   1  &   0  &   1  &   1  &   1  &   0  &   1  &   1  &   1 \\
     1  &   0  &   0  &   0  &   1  &   0  &   0  &   0  &   0  &   1 \\
     0  &   0  &   0  &   1  &   0  &   1  &   0  &   0  &   0  &   0 \\
     1  &   2  &   2  &   1  &   2  &   1  &   2  &   1  &   0  &   0
\end{pmatrix}
\label{matA}
\eeq

From the data above, we can see that individuals $1$ and $5$ could be regarded a priori as the main influencers of the network. In our experiment, we asked the 10 individuals 9 different questions $Q1, Q2, \cdots, Q9$, with a binary output, which we codify as $\pm 1$. Using our model, we could see that by fixing the answer of some of the influencers (i.e. $f_i \neq 0$), we could systematically obtain more than $50 \%$ correctness in predicting the answers of the rest of individuals, reaching in some cases very large percentages. This is shown in Table \ref{tabl}. Interestingly, we observe that in practice  individuals $1, 3$ and $7$ seem to have the strongest influence in the outcomes of the network, whereas $5$ does not seem to have much effect. This seems to indicate that the e.g. the tweets by $7$ have more impact than those by $5$. Or in other words, $5$ may have many tweets, but of low impact. Moreover, we could also see that in many situations, by fixing all the answers except one, we could predict the answer of the remaining individual with  a high degree of accuracy. For instance, in the case of questions Q6 and Q9, we obtained respectively $90 \%$ and $70 \%$ accuracy. 

Several remarks are in order. First, we noticed that depending on type of question, influencers may vary. This comes as no surprise, since different people have the strongest influence on the network in different aspects. Second, even for the very small network that we treated, with limited data and a very simple model Hamiltonian, we were already able to obtain quite decent predictions. This is a proof-of-priniciple of the validity of our idea, and we think therefore that more refined models, including more data and bigger networks, will be able to provide much more accurate predictions. Last but not least, we would like to mention that our predictions are based on the lowest-energy eigenstate of the corresponding model Hamiltonian. It should be possible, however, to explore more states in the low-energy space, even with the possibility of assigning a probability (e.g., thermal). In such an analysis one could give probabilistic predictions, relevant for the study of very large networks.  

\begin{table}[h]
\centering
\begin{tabular}{| c || c | c | c | c | c | c | c | c | c |}
\hline
          & Q1 & Q2 & Q3 & Q4 & Q5 & Q6 & Q7 & Q8 & Q9 \\
\hline
\hline
~Fixed~ & 5, 7 & 1, 3 & 1 & 3, 7 & 7 & 1 & 3, 4 & 3, 7 & 1, 7 \\
\hline
~Score~ & $50 \%$ & $50 \%$ & $67 \%$ & $62.5 \%$ & $56 \%$ & $89 \%$& $62.5 \%$& $62.5 \%$ & $75 \%$\\
\hline
\end{tabular}
\caption{Results on the prediction of outcomes for our \emph{Twitter} experiment. For the 9 questions (Q1 - Q9), we show the individuals for which we set $f_i = \pm 1$ according to their answer, and then the score for the prediction of the rest of the outcomes ($\%$ of answers that are correctly predicted).}
\label{tabl}
\end{table}
 
\section{Conclusions} 
\label{conclusions}

Here we have shown that political forecasting can be mapped to finding the ground state configuration of a classical spin system. This spin system can be a combination of $XY$, Ising and vector Potts models, always with two-spin interactions, magnetic fields, and on arbitrary graphs. By reduction to the Ising model the problem is formally an NP-Hard. We have also shown that it can be recasted as both Higher-order and Quadratic Unconstrained Binary Optimization (HUBO / QUBO) Problems. Finally, we have discussed how to construct appropriate models from social data, and conducted an experiment with ten anonymous \emph{Twitter} volunteers, to whom we asked questions of the type ``A or B".  After identifying the relevant influencers, the constructed models were able to predict a good fraction of the outcomes reliably. The results of this very simple experiment already prove the validity of the ideas presented in this paper. 

The results in this paper open a promising, new, and original manner of predicting social trends and forecasting politics employing quantum systems. The idea can also be applied in other contexts, such as sentiment analysis and fake news identification, offering a new perspective of combinatorial optimization algorithms in these scenarios. In future works it would be interesting to explore the validity of the polling system for (open) election data of previous elections of which the end results are already known. Our work offers also a nice example of cross-disciplinary research between physics and social sciences, aka sociophysics. Importantly, for very large networks, we expect that quantum processors will be able to handle the problem in the short-mid term. We see this as a very promising application of quantum computing in solving practical social problems, which deserves future studies.

\bigskip 
\bigskip

{\bf Acknowledgements.-} We acknowledge the ten volunteers who agreed to participate in our experiment. We also acknowledge relevant discussions with Geza Giedke, Enrique Lizaso, and Samuel Mugel, as well as Ikerbasque, DIPC, and UPV/EHU. Additionally, M.S. acknowledges support from Spanish Government PGC2018-095113-B-I00 (MCIU/AEI/FEDER, UE), Basque Government IT986-16, the projects QMiCS (820505) and OpenSuperQ (820363) of the EU Flagship on Quantum Technologies, as well as from the EU FET Open project Quromorphic (828826). This material is also based upon work supported by the U.S. Department of Energy, Office of Science, Office of Advance Scientific Computing Research (ASCR), under field work proposal number ERKJ333. R. I. acknowledges the support of the Basque Government Ikasiker Grant. 

{\bf Conflict of interest.-} On behalf of all authors, the corresponding author states that there is no conflict of interest. 

{}

\end{document}